\def\version{February 12, 2016}

\documentclass[12pt]{article}
\usepackage[textwidth=500pt,textheight=650pt,centering]{geometry} 
\usepackage{amsmath,amssymb,amsthm,cite}
\usepackage{color}

\newtheorem{theorem}{Theorem}

\title{Renormalisation group analysis of 4D
\\
spin models and self-avoiding walk}

\author{
  Roland Bauerschmidt\thanks{Department of Mathematics,
    Harvard University,
    Cambridge, MA 02138, USA.
    E-mail: {\tt brt@math.harvard.edu}.},\;
  David  Brydges\thanks{Department of Mathematics,
    University of British Columbia,
    Vancouver, BC, Canada V6T 1Z2.
    E-mail: {\tt db5d@math.ubc.ca}, {\tt slade@math.ubc.ca}.}\;
  and Gordon Slade$^\dagger$}

\date\version

\begin{document}






\maketitle 

\begin{abstract}
We give an overview of results on critical phenomena in 4 dimensions, obtained
recently using a rigorous renormalisation group method.
In particular, for the $n$-component $|\varphi|^4$ spin model in dimension 4,
with small coupling constant, we prove
that the susceptibility diverges with a logarithmic correction to the
mean-field behaviour with exponent $(n+2)/(n+8)$.
This result extends rigorously to $n=0$, interpreted as
a supersymmetric version of the model that represents exactly the continuous-time
weakly self-avoiding walk. We also analyse the critical two-point function of the
weakly self-avoiding walk, the specific heat and pressure of the $|\varphi|^4$ model,
as well as scaling limits of the spin field close to the critical point.
\end{abstract}

\newcommand{\N}{\mathbb{N}}
\newcommand{\Z}{\mathbb{Z}}
\newcommand{\Q}{\mathbb{Q}}
\newcommand{\R}{\mathbb{R}}
\newcommand{\C}{\mathbb{C}}
\newcommand{\K}{\mathbb{K}}
\newcommand{\T}{\mathbb{T}}
\newcommand{\Ex}{\mathbb{E}}
\newcommand{\1}{{\bf 1}}
\newcommand{\ddp}[2]{\frac{\partial #1}{\partial #2}}



\section{Introduction and results}\label{sec:models}

\paragraph{$|\varphi|^4$ model}

Our results apply to the $n$-component $|\varphi|^4$ model on the
4-dimensional integer lattice $\Z^d$ with $d=4$.
To define the model, we approximate $\Z^d$ by a discrete torus
$\Lambda = \Lambda_N = \Z^d/L^N \Z^d$ of side length $L^N$
with $L$ fixed (large), and eventually $N\to\infty$.
To define the model and set notation, for coupling constants $g>0$, $\nu,z\in\R$,
a subset $X \subseteq \Lambda$, and a field $\varphi: \Lambda \to \R^n$, set
\begin{equation} \label{e:Vdef}
  V_{g,\nu,z}(\varphi,X) = \sum_{x\in X} \left( \frac12 z \varphi_x \cdot (-\Delta\varphi)_x + \frac12 \nu |\varphi_x|^2
    + \frac14 g |\varphi_x|^4 \right)
  .
\end{equation}
The $|\varphi|^4$ model is then defined as the probability measure
\begin{equation} \label{e:phi4def}
  \frac{1}{Z_{g,\nu,\Lambda}} e^{-V_{g,\nu,1}(\varphi,\Lambda)} \, \prod_{x\in\Lambda} d\varphi_x
  ,
\end{equation}
where $d\varphi_x$ is the Lebesgue measure on $\R^n$
and $Z_{g,\nu,\Lambda}$ is a normalisation constant (the partition   function).
Assuming (for now) existence of the limits,
the two-point function and susceptibility are defined by
\begin{equation}
  G_{g,\nu}(x) = \lim_{N\to\infty} \langle \varphi_0\cdot \varphi_x
  \rangle_{g,\nu,\Lambda_N}, \qquad
  \chi(g,\nu) = \sum_{x\in\Z^d} G_{g,\nu}(x),
\end{equation}
where $\langle \cdot \rangle_{g,\nu,\Lambda}$ is the expectation of \eqref{e:phi4def},
and the pressure and (singular part of the) specific heat are
$p(g,\nu) = \lim_{N\to\infty} \frac{1}{|\Lambda_N|} \log Z_{g,\nu,\Lambda_N}$
and $c_H(g,\nu) = \ddp{^2}{\nu^2} p(g,\nu)$.

\paragraph{Weakly self-avoiding walk}

Let $X$ be a continuous-time simple random walk on $\Z^d$
and denote by $E_0$ the expectation for the process with $X(0)=0 \in \Z^d$.
The \emph{self-intersection local time} up to time $T$ is the random variable
\begin{equation} \label{e:intersection-local-time-def}
  I(T) = \int_0^T \!\! \int_0^T \1_{X(t_1) = X(t_2)} \; dt_1 \, dt_2
  .
\end{equation}
For $g>0$ and $\nu \in \R$, and $x\in \Z^d$, the continuous-time weakly self-avoiding walk
\emph{two-point function} and \emph{susceptibility} are defined by the (possibly infinite) integrals
\begin{equation}
  \label{e:two-point-function-wsaw}
  G_{g,\nu}(x)
  =
  \int_0^\infty
  E_0 \left(
    e^{-g I(T)}
    \1_{X(T)=x} \right)
  e^{- \nu T}
  dT,
  \qquad
  \chi(g,\nu) = \sum_{x\in\Z^d} G_{g,\nu}(x)
  .
\end{equation}
Overviews of results on (weakly) self-avoiding walks can be found in Refs.~\citenum{BDGS12,MS93}.
The weakly self-avoiding walk is believed to be in the same
universality class as the strictly self-avoiding walk.
It is exactly related to a supersymmetric version of the
$|\varphi|^4$ model with a complex bosonic and a complex fermionic field,
and this is the starting point for our analysis\cite{McKa80,PS80,BIS09}.
The fermionic components effectively count negatively,
and we thus refer to weakly self-avoiding walk case as the case $n=0$ of the $|\varphi|^4$ model
(with $0$ interpreted as $2-2$).

\medskip

The following theorem summarises the main results of
Refs.~\citenum{BIS09,Baue13a,BS-rg-norm,BS-rg-loc,BBS-rg-pt,BS-rg-IE,BS-rg-step,BBS-rg-flow,BBS-saw4-log,BBS-saw4,BBS-phi4-log,ST-phi4}.
Here $A \sim B$ stands for $\lim A/B =1$.

\begin{theorem} \label{thm:summary}
Let $d=4$, $n=0,1,2,\dots$, $p>0$, and let $g>0$ be small (depending on $n$ and $p$).
For the $n$-component $|\varphi|^4$ model ($n \ge 1$),
and for the weakly self-avoiding walk ($n=0$),
there exist critical values $\nu_c=\nu_c(g,n)$ such that,
as $\varepsilon \downarrow 0$ respectively $|x|\to\infty$,
the following hold
(with constants $A,B,D>0$ depending on $g,n$, and $C_p$ depending on $p$).
\begin{enumerate}
\item[(i)]
(Ref.~\citenum{BBS-saw4})
For $n \ge 0$, the critical two-point function decays as
  \begin{equation}
    G_{g,\nu_c}(x) \sim B |x|^{-2}.
  \end{equation}
\item[(ii)]
(Refs.~\citenum{BBS-saw4-log,BBS-phi4-log})
For $n \ge 0$, the susceptibility obeys
\begin{align} \label{e:suscept4}
    \chi(g,\nu_c+\varepsilon)
    &\sim A \, \varepsilon^{-1} (\log \varepsilon^{-1})^{(n+2)/(n+8)}.
\end{align}
\item[(iii)]
(Ref.~\citenum{BSTW-clp})
For $n \ge 0$, the correlation length of order $p>0$ obeys
\begin{equation}
  \frac{1}{\chi(g,\nu_c+\varepsilon)}
  \sum_{x} |x|^p G_{g,\nu_c+\varepsilon}(x)
  \sim
  C_p A^{-p/2}
  \varepsilon^{-p/2} (\log \varepsilon^{-1})^{p(n+2)/(2n+16)}
  .
\end{equation}
\item[(iv)]
(Ref.~\citenum{BBS-phi4-log})
For $n \ge 1$, the specific heat obeys
\begin{equation}
  \label{e:cHasy}
  c_H(g,\nu_c+\varepsilon)
  \sim D
  \begin{cases}
    (\log \varepsilon^{-1})^{(4-n)/(n+8)} & (n=1,2, 3)\\
    \log \log \varepsilon^{-1} & (n=4)\\
    1 & (n>4).
  \end{cases}
\end{equation}
\item[(v)]
(Ref.~\citenum{BBS-phi4-log})
For $n \ge 1$,
the spin field on the discrete torus of side length $L^N$
converges weakly to white noise if $\nu>\nu_c$,
and to a massive Gaussian free field if $\nu \downarrow \nu_c$ as $N\to\infty$
appropriately.
\item[(vi)]
(Ref.~\citenum{ST-phi4})
For $n\geq 0$, several multi-point functions
have interesting $n$-dependent logarithmic corrections.
\end{enumerate}
The limits defining the quantities on the left-hand sides are taken along the sequence
$\Lambda_N$ with $L$ large enough, and the statement includes their existence in this case.
For $n=0,1,2$, independence of the sequence of most limits is known by other methods.
\end{theorem}

Item~(iii) was obtained with Tomberg and Wallace, and
(vi) with Tomberg. All results rely on a general renormalisation group
method, outlined in the remainder of these proceedings.
Several cases of the above results have been proved previously by different renormalisation group methods.
In particular,
(i) and a case of (vi) was proved for $n=1$ in Refs.~\citenum{GK85,GK86},
(i) for $n=1$ was independently proved in Ref.~\citenum{FMRS87},
versions of (ii), (iii) for $n=1$ were obtained in Refs.~\citenum{Hara87,HT87},
and (i) for a version of $n=0$ in Ref.~\citenum{IM94}.
A hierarchical version of the 4-dimensional weakly self-avoiding walk was
studied in Refs.~\citenum{BEI89,BI03c,BI03d,GI95},
also for complex $\nu$, which permits inversion of
the Laplace transforms $G_{g,\nu}(x)$ in \eqref{e:two-point-function-wsaw} and
the analysis of the end-to-end distance.
The above critical behaviour
was first predicted over 40 years ago using non-rigorous methods;
see in particular Refs.~\citenum{LK69,WR73,BGZ73}.

\section{Method}

The results of Theorem~\ref{thm:summary} are proved by a rigorous
version of Wilson's renormalisation group \cite{WK74}, developed in
Refs.~\citenum{BIS09,BGM04,Baue13a,BS-rg-norm,BS-rg-loc,BBS-rg-pt,BS-rg-IE,BS-rg-step,BBS-rg-flow}.
This method applies to bosonic fields (standard probability theory), fermionic
fields (Grassmann fields), or both, and is compatible with supersymmetry.
For brevity, we only discuss the (bosonic) $|\varphi|^4$ model.

From now on, we identify $V = (g,\nu,z) \in \R^3$ with the function $V_{g,\nu,z}$ defined in \eqref{e:Vdef}.
Then for $m^2>0$ and $V_0=(g_0,\nu_0,z_0)$ with $z_0 > -1$ and $g_0>0$,
we define
\begin{equation} \label{e:ZNdef}
  Z_N(\varphi) = (\Ex_C\theta Z_0)(\varphi), \quad Z_0 = e^{-V_0(\varphi,\Lambda)}, \quad C=(-\Delta+m^2)^{-1}
\end{equation}
where $\Ex_C\theta F$ denotes the convolution of $F$ with the Gaussian measure with covariance $C$.
By a change of variables, the original model can be studied in terms of $Z_N$ with $g_0 = g(1+z_0)^2$ and $\nu_0 = (1+z_0)\nu-m^2$.
It will be useful to carry out the analysis as a function of the four parameters $(m^2,g_0,\nu_0,z_0)$,
and specialise later.

\paragraph{Progressive integration}
The starting point for the analysis of $Z_N$ is a
positive definite
\emph{finite-range decomposition}\cite{Baue13a,BGM04} of
the operator $(-\Delta+m^2)^{-1}$ ($m^2>0$) on $\Lambda_N$ as
\begin{equation}
  (-\Delta_{\Lambda_N}+m^2)^{-1} = C_1 + \cdots + C_{N-1} + C_{N,N}
  ,
\end{equation}
satisfying $C_{j;x,y} = 0$ if $|x-y| > \frac12 L^j$ (finite range property),
the estimates
$|\nabla^\alpha C_{j;x,y}| = O((1+L^{2(j-1)}m^2)^{-s}L^{-(d-2+|\alpha|_1)(j-1)})$ for any $s>0$ and all $j < N$ (scaling estimates),
and additional less significant properties.
Moreover, similar estimates hold for $C_{N,N}$ for $m^2 \geq c L^{-2(N-1)}$,
and we thus often write $C_N$ instead of $C_{N,N}$.
Such a covariance decomposition enables a progressive evaluation\cite{BCGNOPS78} of $Z_N$ as the last element of
\begin{equation}
  Z_{j+1} = \Ex_{C_{j+1}}\theta Z_j, \quad Z_0 = e^{-V_0(\Lambda)}.
\end{equation}
The torus $\Lambda_N$ is decomposed as the union over $\mathcal{B}_j$ of disjoint blocks
of side length $L^j$ where $\mathcal{B}_j$ is
such that each block $b \in \mathcal{B}_{j}$ is completely contained in a block $B \in \mathcal{B}_{j+1}$.
The set of polymers $\mathcal{P}_j$ consists of unions of blocks in $\mathcal{B}_j$. For any $X \in \mathcal{P}_j$,
we denote by $\mathcal{B}_j(X)$ the blocks contained in $X$.
The finite range property asserts that the restrictions
of a Gaussian field $\zeta$ with covariance $C_j$
to two polymers in $\mathcal{P}_j$ that do not touch are independent.

\paragraph{Renormalisation group}

The renormalisation group map is a description of the \emph{global} map $Z_j\mapsto Z_{j+1}$ in terms of \emph{local} coordinates
$I_j$ and $K_j$, where $I_j$ corresponds to the relevant and marginal directions in the Wilson renormalisation group\cite{WK74},
and $K_j$ to the irrelevant directions. More concretely, there is an explicit function
$W_j$ such that
the coordinate
\begin{equation}
  I_j(X,\varphi) = \prod_{B \in \mathcal{B}_j(X)} e^{-V_j(B,\varphi)}(1+W_j(B, V_j,\varphi)), \quad (X\in \mathcal{P}_j)
\end{equation}
is completely determined by three \emph{coupling constants} $V_j=(g_j,\nu_j,z_j) \in \R^3$, 
and $I_j$ factors over $j$-blocks.
The irrelevant coordinate $K_j(X,\varphi)$ has the weaker factorisation property
\begin{equation}
  K_j(X \cup Y,\varphi) = K_j(X,\varphi)K_j(Y,\varphi) \quad \text{for $X,Y \in \mathcal{P}_j$ that do not touch}.
\end{equation}
Both $I_j(X,\varphi)$ and $K_j(X,\varphi)$ have the locality property that they only depend on $\varphi$ in a neighbourhood of $X$,
as well as the normalisation $I_j(\varnothing) = K_j(\varnothing) = 1$.
They can be multiplied by the \emph{circle product}\cite{BY90}
\begin{equation}
  (I_j \circ K_j)(X,\varphi) = \sum_{Y \in \mathcal{P}_j(X)} I_j(X \setminus Y,\varphi) K_j(Y,\varphi).
\end{equation}
For $j=0$ one then has $Z_j(\varphi) = e^{-u_j|\Lambda|}(I_j \circ K_j)(\Lambda,\varphi)$,
with $u_0=0$, $W_0=0$, and $K_0(X,\varphi) = {\bf 1}_{X = \varnothing}$.
The \emph{renormalisation group map} is a lifting of the map $Z_j \mapsto Z_{j+1}$
to a map $(u_j, I_j, K_j) \mapsto (u_{j+1}, I_{j+1},K_{j+1})$, with $u_j \in \R$, such that
\begin{equation}
  e^{-u_{j}|\Lambda|} \Ex_{C_{j+1}}\theta (I_j \circ K_j)(\Lambda,\varphi) = e^{-u_{j+1}|\Lambda|} (I_{j+1} \circ K_{j+1})(\Lambda,\varphi).
\end{equation}

\paragraph{Flow of coupling constants}

In Ref.~\citenum{BBS-rg-pt}, the map $V_j \mapsto V_{j+1}$ is defined to second order by perturbation theory.
In Refs.~\citenum{BS-rg-IE,BS-rg-step}, the non-perturbative correction and the complete map $(V_j,K_j) \mapsto (V_{j+1},K_{j+1})$
are defined, as well suitable function spaces of $K_j$ and estimates that show that $K_j$ is
contractive in these spaces.

In particular, the evolution of $V_j$ and thus $I_j$ is determined by
a flow of coupling constants, which similarly as in Wilson's non-rigorous analysis, are given by
\begin{align}
  g_{j+1} &= g_j - \beta_j g_j^2 + r_{g,j} \label{e:flowg}\\
  \mu_{j+1} &= L^2\mu_j\left(1-\frac{n+2}{n+8}\beta_j g_j\right) + (\,\cdots) + r_{\mu,j}. \label{e:flowmu}
\end{align}
Here $\mu_j = L^{2j} \nu_j$,
the $(\,\cdots)$ denote other explicit terms which are at most quadratic in $V$,
and the $r$ are non-perturbative remainders that depend on $K_j$ and are third order in $V$.
The explicit flow of $z_j$ is also important, but conceptually less significant, and
we mostly ignore it in this exposition.
The coefficients $\beta_j$ are given by
\begin{equation}
  \beta_j = \sum_{x\in\Z^d} \Big(w_{j+1}(x)^2 -w_j(x)^2\Big), \quad w_j(x) = \sum_{k=1}^jC_k(x).
\end{equation}
To study the approach of the critical point rather than only the critical point itself,
the $\beta_j$ here depend on $m^2>0$ through the covariances $C_k$.
They have asymptotic behaviour $\lim_{m^2\downarrow 0} \beta_j \sim (n+8) (\log L)/(16\pi^2)$ as $j \to\infty$,
and obey $\lim_{N\to\infty} \sum_j \beta_j \to (n+8)B_{m^2}$
where $B_{m^2} = \sum_{x\in \Z^4} [(-\Delta_{\Z^4}+m^2)^{-1}_{0x}]^2
\sim (n+8)\log m^{-2}/(16\pi^2)$ is
the \emph{bubble diagram} of the free Green function. The logarithmic divergence of $B_{m^2}$ is ultimately
responsible for the criticality of $d=4$ and the logarithmic corrections in Theorem~\ref{thm:summary}.

The control of $K_j$ is at the heart of the issues to obtain a mathematically rigorous result.
The analysis in Refs.~\citenum{BS-rg-IE,BS-rg-step} exploits the finite range property of the covariances $C_k$
to avoid the need for cluster expansions.
An example of this approach in a simpler context can be found in Ref.~\citenum{Bryd09}.

The (non-hyperbolic) dynamical system $(V_j,K_j) \mapsto (V_{j+1},K_{j+1})$ is analysed in Refs.~\citenum{BBS-rg-flow,BBS-saw4-log}.
For $(m^2,g_0) \in (0,\delta)^2$ with $\delta>0$ small, initial conditions $(\nu_0,z_0) = (\nu_0^c(m^2,g_0), \nu_0^c(m^2,g_0))$
are determined such that $V_j$ remains bounded and $K_j\to 0$, as $j\to\infty$.
Along this renormalisation group trajectory the observables discussed in Theorem~\ref{thm:summary} are studied.
This will be exemplified in the case of the susceptibility.
The susceptibility is also fundamental to relate $(\nu^c_{0},z^c_{0})$
to the critical points $\nu_c(g)$ of the original models,
using the change of variables mentioned below \eqref{e:ZNdef} and implicit function theory.

\paragraph{Susceptibility}

We sketch the proof of \eqref{e:suscept4}.
For a test function $h: \Lambda \to \R$, set $\Sigma_N(h) = \Ex_C(Z_0(\varphi)e^{(\varphi,h)})$. Then,
by completion of the square,
\begin{equation} \label{e:Sigma}
  \frac{\Sigma_N(h)}{\Sigma_N(0)}
  = e^{\frac12 (h,Ch)} \frac{Z_N(Ch)}{Z_N(0)}
  = e^{\frac12 (h,Ch)} \frac{I_N(\Lambda,Ch)+K_N(\Lambda, Ch)}{I_N(\Lambda,0)+K_N(\Lambda,0)},
\end{equation}
using that $I_N \circ K_N(\Lambda) = I_N(\Lambda)+K_N(\Lambda)$ since there is only one $N$-block on $\Lambda_N$.
Assuming that $(g,\nu)$ and $(m^2,g_0,\nu_0,z_0)$ are related as below \eqref{e:ZNdef},
the susceptibility is obtained (up to a factor $(1+z_0)^2$)
by differentiating twice with respect to a constant test function $h$.
In particular, if $(m^2,g_0,\nu_0,z_0)=(m^2,g_0,\nu_0^c,z_0^c)$ is critical according to the dynamical system analysis,
then $K_N \to 0$ in a suitable norm,
and using $C1 = m^{-2}1$ for constant test function $1_x=1$ as well as the explicit form of $I_N$,
we obtain the \emph{identity}
\begin{align} \label{e:chi}
  \chi(g,\nu)
  &= (1+z_0) \lim_{N\to\infty} \left( \frac{1}{m^2} -
    \frac{\nu_N}{m^{4}} +  \frac{1}{m^4} \frac{1}{|\Lambda|} (D^2W(\Lambda;0;1,1)+D^2K(\Lambda;0;1,1)) \right)
    \nonumber\\
  &= \frac{1+z_0}{m^2}
  .
\end{align}
In particular, $\nu \downarrow \nu_c(g)$ corresponds to $m^2 \downarrow 0$ under the critical choice
of the four coupling constants, and the singular behaviour
of $\chi$ at $\nu=\nu_c(g)$ is encoded in the relationship between $m^2$ and $(g,\nu)$.
To understand $\chi$,
we derive an equation for $\ddp{}{\nu} \chi = (1+z_0) \ddp{}{\nu_0} \chi$.
The derivative can be taken inside the limit in \eqref{e:chi},
and is taken with $m^2,g_0,z_0$ fixed. Then the $\nu_0$-derivative of $1/m^2$
vanishes and the main contribution to $\ddp{}{\nu}\chi$ is given by $-\nu_N'/m^4$
(with the contribution due to $K_N$ again subleading), where the prime
denotes the derivative with respect to $\nu_0$.
By differentiating \eqref{e:flowg}--\eqref{e:flowmu},
along the critical trajectory, for which coupling constants are controlled,
it can be shown that
\begin{equation} \label{e:muprime}
  \nu_j' \sim (1+O(g)) \left(\frac{g_j}{g_0}\right)^{(n+2)/(n+8)}.
\end{equation}
The coupling constant $g_j$ tends to an $m^2$-dependent limit $g_\infty$.
As $m^2\downarrow 0$,
\begin{equation} \label{e:ginfty}
  g_\infty \sim \frac{1}{(n+8)B_{m^2}} \sim \frac{16\pi^2}{(n+8)\log m^{-2}}
  .
\end{equation}
This leads to
\begin{align} \label{e:dchi}
  \ddp{\chi}{\nu}(g,\nu)
  &= \frac{(1+z_0)^2}{m^4} \lim_{N\to\infty}
  \left( - \nu_N' + \ddp{}{\nu_0} \frac{1}{|\Lambda|} ( D^2W(\Lambda;0;1,1) +D^2K(\Lambda;0;1,1) )\right)
  \nonumber\\
  &\sim c\frac{(\log m^{-2})^{(n+2)/(n+8)}}{m^4}
  .
\end{align}
From \eqref{e:chi} and \eqref{e:dchi} we obtain $\ddp{}{\nu}\chi \sim c (\log \chi)^{(n+2)/(n+8)} \chi^2$ as $m^2\downarrow 0$,
and the claim
\begin{equation}
  \chi(g,\nu_c(g)+\varepsilon) \sim A \varepsilon^{-1} (\log \varepsilon^{-1})^{(n+2)/(n+8)} \quad (\varepsilon \downarrow 0),
\end{equation}
follows.

\paragraph{Other observables}

The analysis of the specific heat follows a similar strategy as that
for the susceptibility.
The pointwise analysis of the two-point and multi-point functions require the analysis of an additional flow of \emph{observable coupling constants},
which depends on the bulk flow \eqref{e:flowg}--\eqref{e:flowmu}, but not vice-versa.
In particular, it is also shown that $G_{\nu_c}(x) \sim (1+z_0) (-\Delta)^{-1}_{0x}$ as $|x|\to\infty$.
Together with \eqref{e:chi} this allows to characterise $m^2$
as the \emph{renormalised mass} 
and $1+z_0$ as the \emph{field strength renormalisation}.
The scaling limit result is obtained by analysing \eqref{e:Sigma} with
general smooth test functions $h$.

\section*{Acknowledgments}

This work on which this article is based was supported in part by NSERC of Canada and
by the U.S.\ NSF under agreement DMS-1128155.

\bibliographystyle{plain}
\bibliography{../rg}

\begin{thebibliography}{10}

\bibitem{Baue13a}
R.~Bauerschmidt.
\newblock A simple method for finite range decomposition of quadratic forms and
  {Gaussian} fields.
\newblock {\em Probab. Theory Related Fields}, {\bf 157}:817--845, (2013).

\bibitem{BBS-phi4-log}
R.~Bauerschmidt, D.C. Brydges, and G.~Slade.
\newblock Scaling limits and critical behaviour of the $4$-dimensional
  $n$-component $|\varphi|^4$ spin model.
\newblock {\em J. Stat. Phys}, {\bf 157}:692--742, (2014).

\bibitem{BBS-saw4}
R.~Bauerschmidt, D.C. Brydges, and G.~Slade.
\newblock Critical two-point function of the 4-dimensional weakly self-avoiding
  walk.
\newblock {\em Commun.\ Math.\ Phys.}, {\bf 338}:169--193, (2015).

\bibitem{BBS-saw4-log}
R.~Bauerschmidt, D.C. Brydges, and G.~Slade.
\newblock Logarithmic correction for the susceptibility of the 4-dimensional
  weakly self-avoiding walk: a renormalisation group analysis.
\newblock {\em Commun.\ Math.\ Phys.}, {\bf 337}:817--877, (2015).

\bibitem{BBS-rg-pt}
R.~Bauerschmidt, D.C. Brydges, and G.~Slade.
\newblock A renormalisation group method. {III}. {Perturbative} analysis.
\newblock {\em J. Stat. Phys}, {\bf 159}:492--529, (2015).

\bibitem{BBS-rg-flow}
R.~Bauerschmidt, D.C. Brydges, and G.~Slade.
\newblock Structural stability of a dynamical system near a non-hyperbolic
  fixed point.
\newblock {\em Ann. Henri Poincar\'e}, {\bf 16}:1033--1065, (2015).

\bibitem{BDGS12}
R.~Bauerschmidt, H.~Duminil-Copin, J.~Goodman, and G.~Slade.
\newblock Lectures on self-avoiding walks.
\newblock In D.~Ellwood, C.~Newman, V.~Sidoravicius, and W.~Werner, editors,
  {\em Probability and Statistical Physics in Two and More Dimensions}, pages
  395--467. Clay Mathematics Proceedings, vol. 15, Amer. Math. Soc.,
  Providence, RI, (2012).

\bibitem{BSTW-clp}
R.~Bauerschmidt, G.~Slade, A.~Tomberg, and B.~Wallace.
\newblock Finite-order correlation length for 4-dimensional weakly
  self-avoiding walk and $|\varphi|^4$ spins.
\newblock Preprint, (2015).

\bibitem{BCGNOPS78}
G.~Benfatto, M.~Cassandro, G.~Gallavotti, F.~Nicol\`{o}, E.~Oliveri,
  E.~Presutti, and E.~Scacciatelli.
\newblock Some probabilistic techniques in field theory.
\newblock {\em Commun. Math. Phys.}, {\bf 59}:143--166, (1978).

\bibitem{BGZ73}
E.~Br\'ezin, J.C. Le~Guillou, and J.~Zinn-Justin.
\newblock Approach to scaling in renormalized perturbation theory.
\newblock {\em Phys.\ Rev.\ D}, {\bf 8}:2418--2430, (1973).

\bibitem{BEI89}
D.~Brydges, S.N. Evans, and J.Z. Imbrie.
\newblock Self-avoiding walk on a hierarchical lattice in four dimensions.
\newblock {\em Ann. Probab.}, {\bf 20}:82--124, (1992).

\bibitem{Bryd09}
D.C. Brydges.
\newblock Lectures on the renormalisation group.
\newblock In S.~Sheffield and T.~Spencer, editors, {\em Statistical Mechanics},
  pages 7--93. American Mathematical Society, Providence, (2009).
\newblock IAS/Park City Mathematics Series, Volume 16.

\bibitem{BGM04}
D.C. Brydges, G.~Guadagni, and P.K. Mitter.
\newblock Finite range decomposition of {Gaussian} processes.
\newblock {\em J. Stat. Phys.}, {\bf 115}:415--449, (2004).

\bibitem{BI03c}
D.C. Brydges and J.Z. Imbrie.
\newblock End-to-end distance from the {G}reen's function for a hierarchical
  self-avoiding walk in four dimensions.
\newblock {\em Commun. Math. Phys.}, {\bf 239}:523--547, (2003).

\bibitem{BI03d}
D.C. Brydges and J.Z. Imbrie.
\newblock {G}reen's function for a hierarchical self-avoiding walk in four
  dimensions.
\newblock {\em Commun. Math. Phys.}, {\bf 239}:549--584, (2003).

\bibitem{BIS09}
D.C. Brydges, J.Z. Imbrie, and G.~Slade.
\newblock Functional integral representations for self-avoiding walk.
\newblock {\em Probab.\ Surveys}, {\bf 6}:34--61, (2009).

\bibitem{BS-rg-norm}
D.C. Brydges and G.~Slade.
\newblock A renormalisation group method. {I}. {Gaussian} integration and
  normed algebras.
\newblock {\em J. Stat. Phys}, {\bf 159}:421--460, (2015).

\bibitem{BS-rg-loc}
D.C. Brydges and G.~Slade.
\newblock A renormalisation group method. {II}. {Approximation by local
  polynomials}.
\newblock {\em J. Stat. Phys}, {\bf 159}:461--491, (2015).

\bibitem{BS-rg-IE}
D.C. Brydges and G.~Slade.
\newblock A renormalisation group method. {IV}. {Stability} analysis.
\newblock {\em J. Stat. Phys}, {\bf 159}:530--588, (2015).

\bibitem{BS-rg-step}
D.C. Brydges and G.~Slade.
\newblock A renormalisation group method. {V}. {A} single renormalisation group
  step.
\newblock {\em J. Stat. Phys}, {\bf 159}:589--667, (2015).

\bibitem{BY90}
D.C. Brydges and H.-T. Yau.
\newblock Grad $\phi$ perturbations of massless {Gaussian} fields.
\newblock {\em Commun. Math. Phys.}, {\bf 129}:351--392, (1990).

\bibitem{FMRS87}
J.~Feldman, J.~Magnen, V.~Rivasseau, and R.~S\'en\'eor.
\newblock Construction and {Borel} summability of infrared {$\Phi^4_4$} by a
  phase space expansion.
\newblock {\em Commun. Math. Phys.}, {\bf 109}:437--480, (1987).

\bibitem{GK85}
K.~Gaw\c{e}dzki and A.~Kupiainen.
\newblock Massless lattice $\varphi^4_4$ theory: Rigorous control of a
  renormalizable asymptotically free model.
\newblock {\em Commun. Math. Phys.}, {\bf 99}:199--252, (1985).

\bibitem{GK86}
K.~Gaw\c{e}dzki and A.~Kupiainen.
\newblock Asymptotic freedom beyond perturbation theory.
\newblock In K.~Osterwalder and R.~Stora, editors, {\em Critical Phenomena,
  Random Systems, Gauge Theories}, Amsterdam, (1986). North-Holland.
\newblock Les Houches 1984.

\bibitem{GI95}
S.E. Golowich and J.Z. Imbrie.
\newblock The broken supersymmetry phase of a self-avoiding random walk.
\newblock {\em Commun.\ Math.\ Phys.}, {\bf 168}:265--319, (1995).

\bibitem{Hara87}
T.~Hara.
\newblock A rigorous control of logarithmic corrections in four dimensional
  $\varphi^4$ spin systems. {I}. {Trajectory} of effective {Hamiltonians}.
\newblock {\em J. Stat. Phys.}, {\bf 47}:57--98, (1987).

\bibitem{HT87}
T.~Hara and H.~Tasaki.
\newblock A rigorous control of logarithmic corrections in four dimensional
  $\varphi^4$ spin systems. {II}. {Critical} behaviour of susceptibility and
  correlation length.
\newblock {\em J. Stat. Phys.}, {\bf 47}:99--121, (1987).

\bibitem{IM94}
D.~Iagolnitzer and J.~Magnen.
\newblock Polymers in a weak random potential in dimension four: rigorous
  renormalization group analysis.
\newblock {\em Commun. Math. Phys.}, {\bf 162}:85--121, (1994).

\bibitem{LK69}
A.I. Larkin and D.E. Khmel'Nitski\u{i}.
\newblock Phase transition in uniaxial ferroelectrics.
\newblock {\em Soviet Physics JETP}, {\bf 29}:1123--1128, (1969).
\newblock {English} translation of {\it Zh.\ Eksp.\ Teor.\ Fiz.} {\bf 56},
  2087--2098, (1969).

\bibitem{MS93}
N.~Madras and G.~Slade.
\newblock {\em The Self-Avoiding Walk}.
\newblock Birkh{\"a}user, Boston, (1993).

\bibitem{McKa80}
A.J. McKane.
\newblock Reformulation of $n \to 0$ models using anticommuting scalar fields.
\newblock {\em Phys. Lett. A}, {\bf 76}:22--24, (1980).

\bibitem{PS80}
G.~Parisi and N.~Sourlas.
\newblock Self-avoiding walk and supersymmetry.
\newblock {\em J. Phys. Lett.}, {\bf 41}:L403--L406, (1980).

\bibitem{ST-phi4}
G.~Slade and A.~Tomberg.
\newblock Critical correlation functions for the $4$-dimensional weakly
  self-avoiding walk and $n$-component $|\varphi|^4$ model.
\newblock To appear in \emph{Commun. Math. Phys.}

\bibitem{WR73}
F.J. Wegner and E.K. Riedel.
\newblock Logarithmic corrections to the molecular-field behavior of critical
  and tricritical systems.
\newblock {\em Phys. Rev. B}, {\bf 7}:248--256, (1973).

\bibitem{WK74}
K.G. Wilson and J.~Kogut.
\newblock The renormalization group and the $\epsilon$ expansion.
\newblock {\em Phys. Rep.}, {\bf 12}:75--200, (1974).

\end{thebibliography}

\end{document}